# Build a training interface to install the bat's echolocation skills in humans

Miyoko TSUMAKI[1,a], Yu TESHIMA[1], Takao TSUCHIYA[2], Kaoru ASHIHARA[3],
Kohta I. KOBAYASI[1], Shizuko HIRYU[1,b]

[1] Graduate School of Life and Medical Sciences, Doshisha University, Japan

[2] Faculty of Science and Engineering, Doshisha University, Japan

[3] Human Informatics and Interaction Research Institute, National Institute of Advanced Industrial Science and Technology (AIST), Japan

## ABSTRACT

Bats use a sophisticated ultrasonic sensing method called echolocation to recognize the environment. Recently, it has been reported that normal sight participants with no prior experience in echolocation can improve their ability to perceive the spatial layout of various environments through training to hear echoes (Norman, et al., 2021). In this study, we developed a new training system for human echolocation using the eye tracker. Binaural echoes of consecutive downward linear FM pulses that were inspired by feeding strategies of echolocating bats were simulated using the wave equation finite difference time domain method. In the experiment, the sighted participants were asked to identify the shapes of the hidden target, and the virtual echoes were presented in response to their eye movements on the monitor. Our analyses suggest that the training shortened the sensing time required for shape identification, and the accuracy of shape identification was improved by the strategy to focusing on the edges of the target. The proposed system is considered useful for quantitative verification of human echolocation.

## 1. INTRODUCTION

Blind echolocation experts (EEs) use self-generated mouth clicks to perceive their surroundings via reflected sound waves(echoes) [1]. Meanwhile, it has been reported that even sighted participants with no prior experience in echolocation can improve their spatial recognition skills through training [2]. However, the actual echo measurement requires time and effort for experimentation [3], and there are many unclear points about effective training methods and learning processes of echolocation skills.

In this study, we propose a system that trains participants to learn echolocation skills with generated echoes from virtual targets of various shapes using 3D acoustic simulation. The participants echolocate the target by hearing the virtual echoes that are generated based on their eye movement on the monitor. Using that system, we analyzed the sensing timing, direction, and shape-identification process of the participants based on the gaze position obtained by the eye-tracker, and examined the sensing strategy of the echolocation inexperienced sighted participants.

## 2. METHODS

### 2.1 Estimated Echoes based on the Impulse Responses

Using a three-dimensional acoustic simulation based on a wave equation finite difference time domain (WE-FDTD) method [4], the shape of a dummy head (HATS; Head and Torso Simulator Brüel

[a]mtsumaki2021@gmail.com
[b]shiryu@mail.doshisha.ac.jp

& Kjær TYPE 4128-C) was digitized and saved as the stereolithography (STL) file. The STL models of the HATS and a virtual target were placed in the simulation space (1.5 × 1.5 × 1.5 m) (Fig. 1). The spatial resolutions of the simulations were 1.5 mm, Courant-Friedrichs-Lewy (CFL) =0.5, and the sound velocity was set to 344 m/s which satisfied the condition for 3D simulation stability. A transmitted signal (impulse) was emitted from the mouth of the HATS, and the echo from the virtual target at the entrance of the left and right ear canals of the HATS as the receiving point was calculated based on numerical analyses. This calculated echo was defined as the echo impulse response. At this time, the target was divided into 5 x 5 cm meshes, and the x and z coordinates of the position of the sound source (mouth of the HATS) were set to the x and z coordinates of the center position of each mesh to calculate the virtual echo. There were 8 types of virtual targets for the training task, and together with 5 untrained versions, the test task was performed on 13 different targets (Fig. 2).

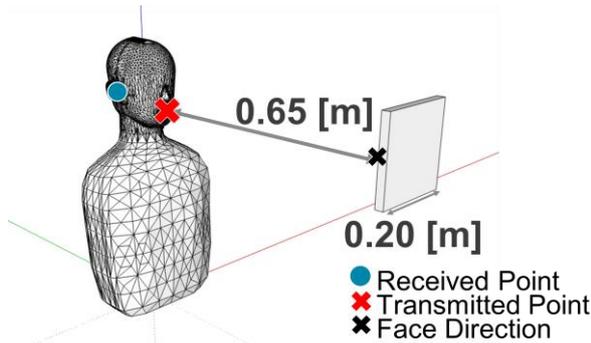

Fig. 1 An example of geometrical configuration of the HATS and the virtual target in a 3D simulation space. The transmitted signal was incident at any position on the target by the HATS tilted to change the direction of the mouth.

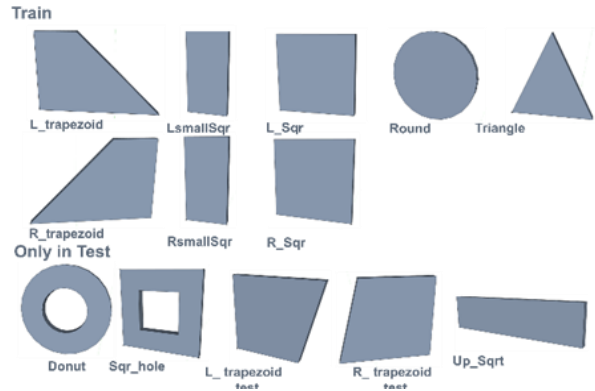

Fig. 2 Virtual targets used in training sessions and test sessions.

## 2.2 Psychoacoustic Experiment

The psychoacoustic experiment was conducted in a sound-attenuated chamber (2.3×1.6×1.4 m). The participants were 1 male and 4 females (22 - 25 years old) with normal sight. It was confirmed that the hearing levels of all participants were within 30 dB in the 0.25 to 3 kHz range based on standard pure-tone audiometry using an audiometer (AA-77A, RION, Tokyo, Japan). The psychoacoustic experiment using humans was approved by the ethics committee of Doshisha University. After obtaining informed consent from the participants and conducting a hearing test, we began the psychoacoustic experiment. One sound stimulus consisted of, a 10 ms downward linear FM sound (7 kHz - 1 kHz) that was convoluted with the echo impulse response described 2.1, and 30 repeated sounds were produced at 30 ms intervals to simulate the buzz sound inspired by feeding strategies of echolocating bats. The presentation of the sound stimuli to the participants was controlled by PsychoPy3 (Standalone 2021.2.3 for 64bit Windows (using Python3.6)) [5], and the presentation of the sound stimuli to the participants was controlled by PsychoPy3, and the sound stimuli were presented via headphones (MDR-CD900ST; SONY) through the audio interface (OCTA-CAPTURE UA–1010; Roland) from the personal computer.

The participants' gaze position (point of gaze, POG) was recorded using a 150Hz desktop eye-tracker (GP3HD Eye tracker, Gazepoint Vancouver, Canada). The mesh-like guideline was displayed on the monitor, and virtual echoes from the mesh corresponding to the POG were presented from headphones. Participants freely sensed the shape of a virtual target hidden on the monitor with their gaze voluntarily. When the participants judged that they could draw the shape, they ended the session by themselves and then drew the shape of the target they recognized on the paper with the mesh on it. There were 8 training trials (with correct feedback) and 7 testing trials (without correct feedback). After completing the trial, a written questionnaire was conducted.

## 3. RESULTS and DISSCUTION

### 3.1 Participant's sensing strategy in the virtual echolocation training

Figure 3 shows a result of the shape of the target (shown in red dashed line) drawn by one participant after sensing. The degree of similarity is calculated as follows. To quantitatively evaluate the difference between the shape drawn by the participant and the outline of the virtual target presented (hereinafter called "difference"), we calculate it by using matchShapes function of the OpenCV library based on the Hu moment [6]. If the threshold was less than 0.02, it was judged to be the same shape (e.g., The lower the result, the better the match, and the closer to zero the returned value, the higher the similarity). The difference in the trained targets of all participants was 0.11 on average, and the average of the untrained (i.e., try for the first time) targets was 0.17, which suggests that shape discrimination to some extent even for the untrained target. In addition, the average sensing time for the training session until shape identification was 179.6 seconds. The average sensing time for the test session (trained target) was 155.0 seconds and the average sensing time for the test session (untrained target) was 167.0 seconds, which was shorter than the training session. From this result, it is considered that there is a learning effect of echolocation by training.

The POG data were analyzed for differences in the distribution among trained targets and untrained targets. Comparing the training session and the test session (trained target), the test session (trained target) tended to gaze more at the edges and outside of the target, especially in participants who showed results with low differences in shapes. In addition, more attention to the edge and shorter sensing times were observed for trained targets than for untrained targets in the test session. These results suggest that the participants modified their sensing strategies during the training process to fixate their gaze on the edges or outside of the target. The POG distribution of each participant and the responses to the questionnaire after all sessions indicated that the virtual echoes from the target edges were distinctive.

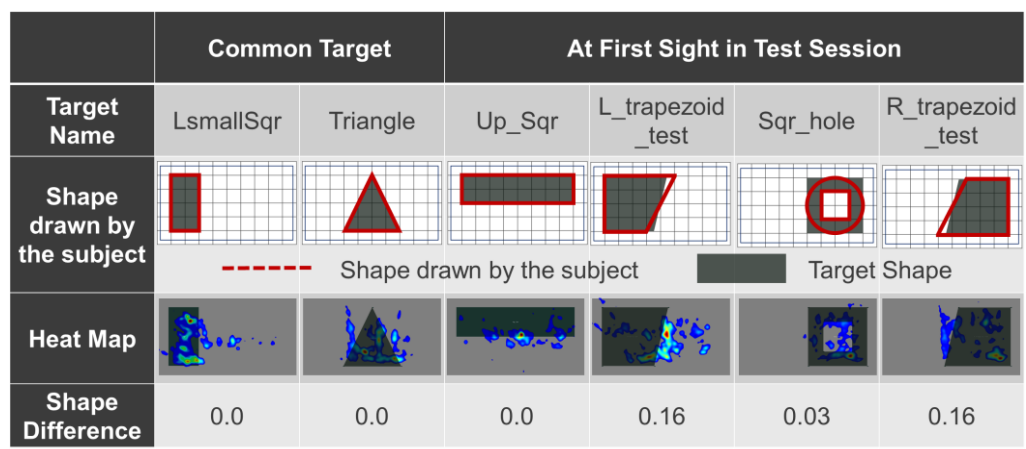

Fig. 3 Shapes drawn in questionnaires(top), heatmaps in the test sessions (middle) and difference values between the target and the shape drawn by the subject (bottom). The red dashed line indicates the shapes drawn by the subject.

## 4. CONCLUSION

In this study, we propose a system that participants listen to virtual echoes while freely sensing by the gaze of sight and investigated the learning process of human echolocation. The results show that the accuracy of shape identification using echolocation is improved by participants changing to sensing strategies that focus on the edge of the virtual targets. This echolocation training system is an effective quantitative approach to the echolocation learning process.